\documentclass{elsarticle}

\usepackage{lineno,hyperref}
\modulolinenumbers[5]
\usepackage{graphicx} 
\graphicspath{ {./pictures/} }
\usepackage{float}
\usepackage{multirow}

\usepackage{xcolor}

\usepackage{subfigure}

\journal{Optics \& Laser Technology}

\begin{document}

\begin{frontmatter}

\title{Laser spot measurement using cost-affordable devices}

\author[pucp]{M. ~Bonnett Del Alamo}
\author[pucp]{C. ~Soncco}
\author[pucp]{R. ~Helaconde}
\author[pucp]{J. ~Bazo}
\author[pucp]{A.M. ~Gago}

\address[pucp]{Secci\'on F\'isica, Departamento de Ciencias, Pontificia Universidad Cat\'olica del Per\'u, Av. Universitaria 1801, Lima 32, Per\'u}

\begin{abstract}

We have designed and tested an automated simple setup for quickly measuring the profile and spot size of a Gaussian laser beam using three cost-affordable light sensors. Two profiling techniques were implemented: imaging for the CMOS 2D array (webcam) and scanning knife-edge-like using a single photodiode and an LDR. The methods and sensors were compared to determine their accuracy using lasers of two different wavelengths and technologies. 

We verify that it is possible to use a low-cost webcam to determine the profile of a laser with 1\% uncertainty on the beam waist, 1.5\% error on the waistline position, and less than 3\% error in determining the minimum spot radius. The photodiode measurement is the most stable since it is not affected by the change in laser intensity. In addition, we show that it is possible to use an inexpensive LDR sensor to estimate the laser spot size with an 11\% error.

\end{abstract}


\end{frontmatter}


\section{Introduction}

A precise characterization of the laser profile and the measurement of its spot size is needed for several scientific, industrial, medical, and instrumental applications \cite{atsumi1994medical, dickey2014laser}. For example, this kind of characterization is important in Laser Induced Breakdown Spectroscopy (LIBS), an atomic emission spectroscopy where the laser focuses to form a plasma, which atomizes samples \cite{Laser_beam_LIBS} or in Laser Tissue Soldering (LTS), where a protein solution is thermally denatured and cross-linked to obtain a strong bond between tissues or tissue and a wound dressing \cite{Martin_Betz_LTS}. 
Another application is related to testing new silicon pixel sensor technology \cite{AglieriRinella:2017lym}, where it is desired to activate individual pixels using a narrow and focused laser beam. The laser profile has to be known in advance to avoid clustering (i.e. illuminating several pixels at the same time). Several methods for measuring a Gaussian laser diameter have been developed, for a summary see \cite{NG20071098}.

There are studies for this purpose that use a CMOS-based camera \cite{Purvis20191977,Hossain20155156} and others based on a quadrant photodiode \cite{NG20071098, LU20143519, Hermosa:11}, which is generally used for measuring the position of the beam. There are also alternative methods \cite{Mylonakis20189863, Cherri2003, Cherri2010, araujo2009} for estimating the diameter of a Gaussian laser beam.

The aim of this work is to offer cost-affordable, automated precise methods to measure, using a simple experimental setup, the spot profile of a Gaussian laser beam. This setup uses a Raspberry Pi (small single-board computer). We are also able to find the smallest spot size in the focal plane, known as beam waist. 

Three different light sensors (a CMOS webcam, a single photodiode and a LDR (Light Dependent Resistor) were used and compared to measure the radius of the laser beam. The performance of the methods and sensors were benchmarked against lasers of two different wavelengths and technologies. In addition, we will show that it is possible to use an LDR sensor, which is very cost-affordable, to roughly estimate the spot size.

This paper is divided as follows: we first describe in Sec. \ref{sec:laser_modelling} the properties and parameters of a Gaussian laser beam. Then, in Sec. \ref{sec:expt_setup} we outline the general experimental setup and the specific used light sensors and lasers. In Sec. \ref{Laser beam radius measurement} we give the results of the laser radius and waistline measurements, as well as the comparison/cross-check between different light sensors. Finally, in Sec. \ref{sec:discussion} we discuss our findings and give our conclusions.

\section{Laser beam modelling}
\label{sec:laser_modelling}

There are several types of laser profiles including Gaussian, multi-mode, tilted, flat top and irregular beams \cite{dickey2014laser}. A laser with a Gaussian irradiance profile can be easily modeled with well-known mathematical expressions. Thus we select this type of beam for testing the performance of the different sensors. 

\begin{figure}[H]
\centering
\includegraphics[width=7cm]{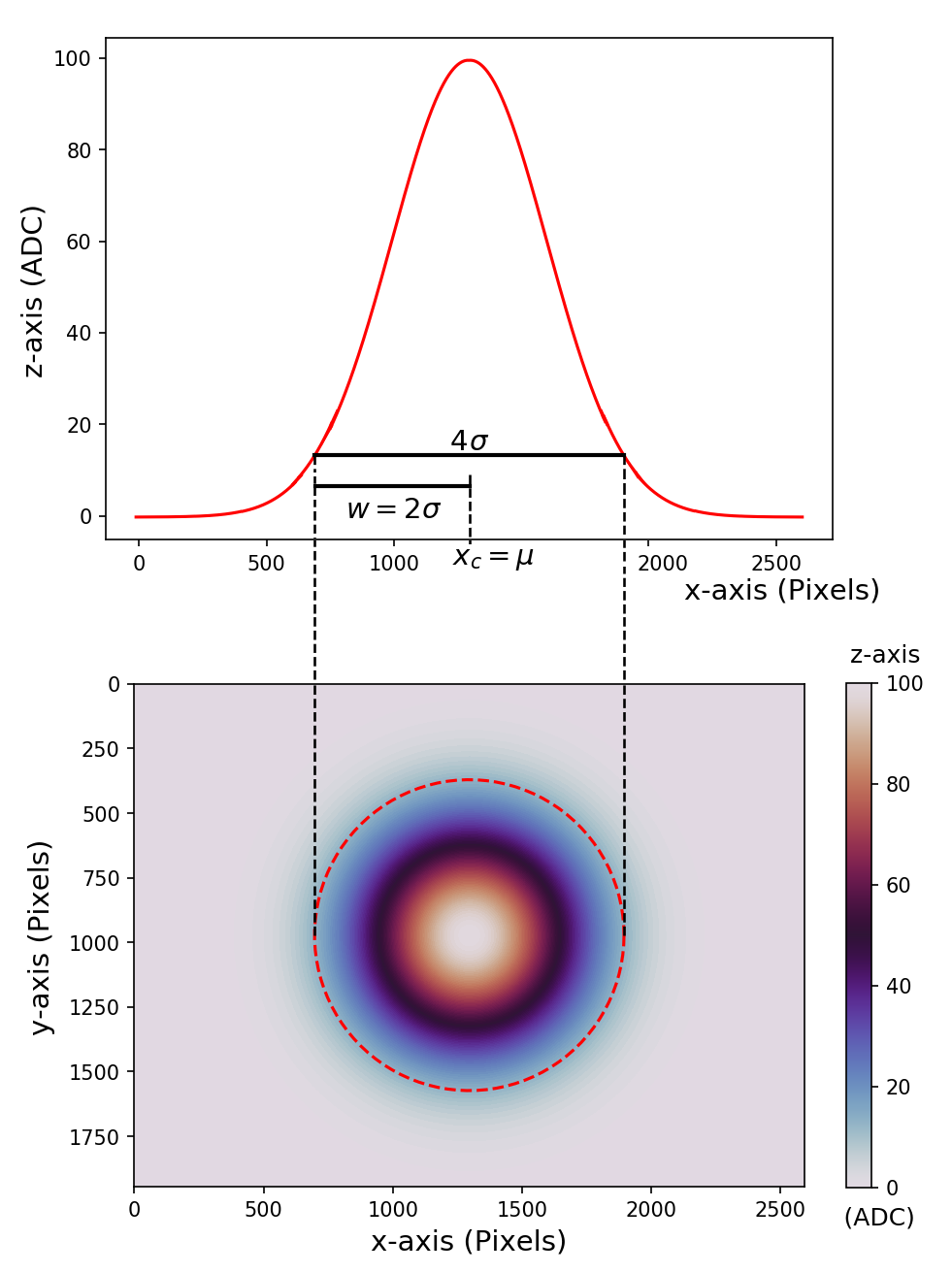}
\caption{Top: projection in the XZ plane of a Gaussian beam, which has a waist ($w$) of $2\sigma$. Bottom: 2D histogram of the Gaussian beam.} 
\label{fig:waistbeam}
\end{figure}

An ideal Gaussian laser beam has a symmetrical irradiation profile around its center, decreasing as the distance from the center increases. The waist of a Gaussian beam, $w$  \cite{Svelto:1338764}, also called radius or spot size, is defined as half the distance across the center of the beam where the irradiance is $1/e^2$ ($\approx$ $13.5$\%) of its maximum. This value is approximately $2\sigma$ of the Gaussian fit, see Fig. \ref{fig:waistbeam}. 

\begin{figure}[htb]
\centering
\includegraphics[width=9cm]{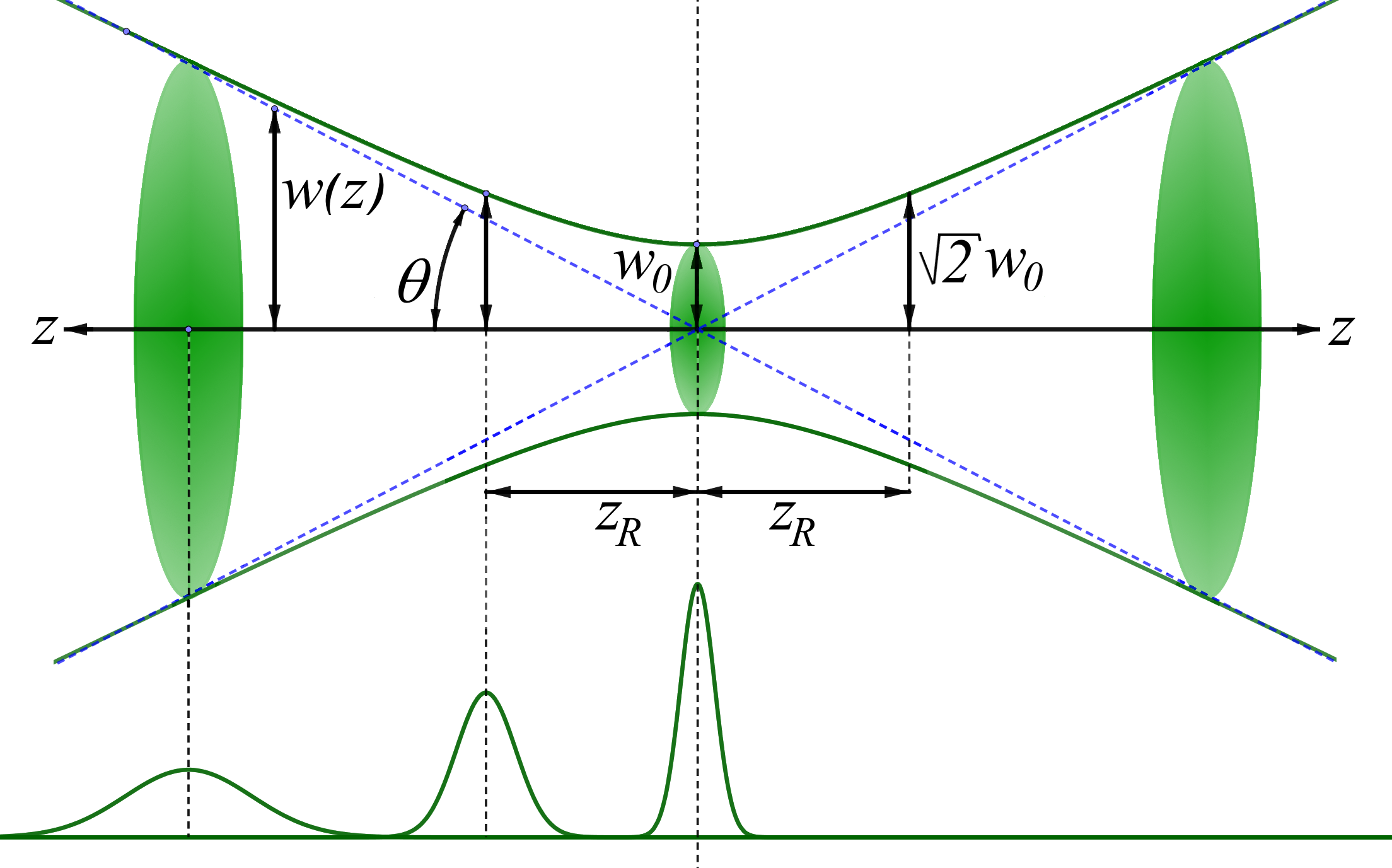}
\caption{Beam profile parameters. Evolution of the radius of the Gaussian beam along the beam $z$-axis as a function of the beam waist ($w_0$), the Rayleigh range ($z_R$) and the divergence angle ($\theta_{d}$). The lower part shows projections at three different distances of a Gaussian beam. The projection at $w_0$ has the narrowest waist.} 
\label{fig:beam_profile}
\end{figure}

The waist at a distance $z$ is given by:
\begin{equation}
w(z) = w_0  \sqrt{ 1+ {\left( \frac{z-z_0}{z_{R}} \right)}^2 }
\label{eq_beam_radius}
\label{eq:w_z}
\end{equation}
where $w_0$ is the radius of the minimum possible cross section (beam minimum waist), $z_0$ is the position of w$_0$ and
$z_R$ is the Rayleigh range, which is defined as the distance from the beam waist to the point where said waist is multiplied by $\sqrt {2}$. A scheme of these parameters is shown in Fig. \ref{fig:beam_profile}. The Rayleigh range is determined by:
\begin{equation}
z_{R}={\frac {\pi w_{0}^{2}n}{\lambda}},
\label{eq:z_R}
\end{equation}
where $\lambda$ is the free-space wavelength and $n$ is the refractive index of the medium where the beam propagates through.

At large distances ($z\gg z_{R}$), $w$ increases linearly with $z$, hence the {\it beam divergence} \cite{Svelto:1338764}, $\theta_{d}$, as seen in Fig. \ref{fig:beam_profile}, given in radians, can be define, due to diffraction, as: 
\begin{equation}
{\theta_{d} =\lim _{z\rightarrow \infty }\arctan {\Big (}{\frac {w(z)}{z}}{\Big )}\simeq {\frac {\lambda }{\pi nw_{0}}} =\frac {w_0}{z_R}}.
\label{eq:divrg}
\end{equation}

For this study, the beam radius equation (Eq. \ref{eq_beam_radius}) will be used to determine the beam profile. To achieve this, it is necessary to measure the values of the laser spot size as a function of the distance to the sensor, as it moves away from the laser. From these data, it is possible to determine the values of: $w_0$, which presents zero divergence, $z_R$ and $z_0$.

\section{Experimental setup}
\label{sec:expt_setup}

\begin{figure}[H]
\centering
\includegraphics[width=10cm]{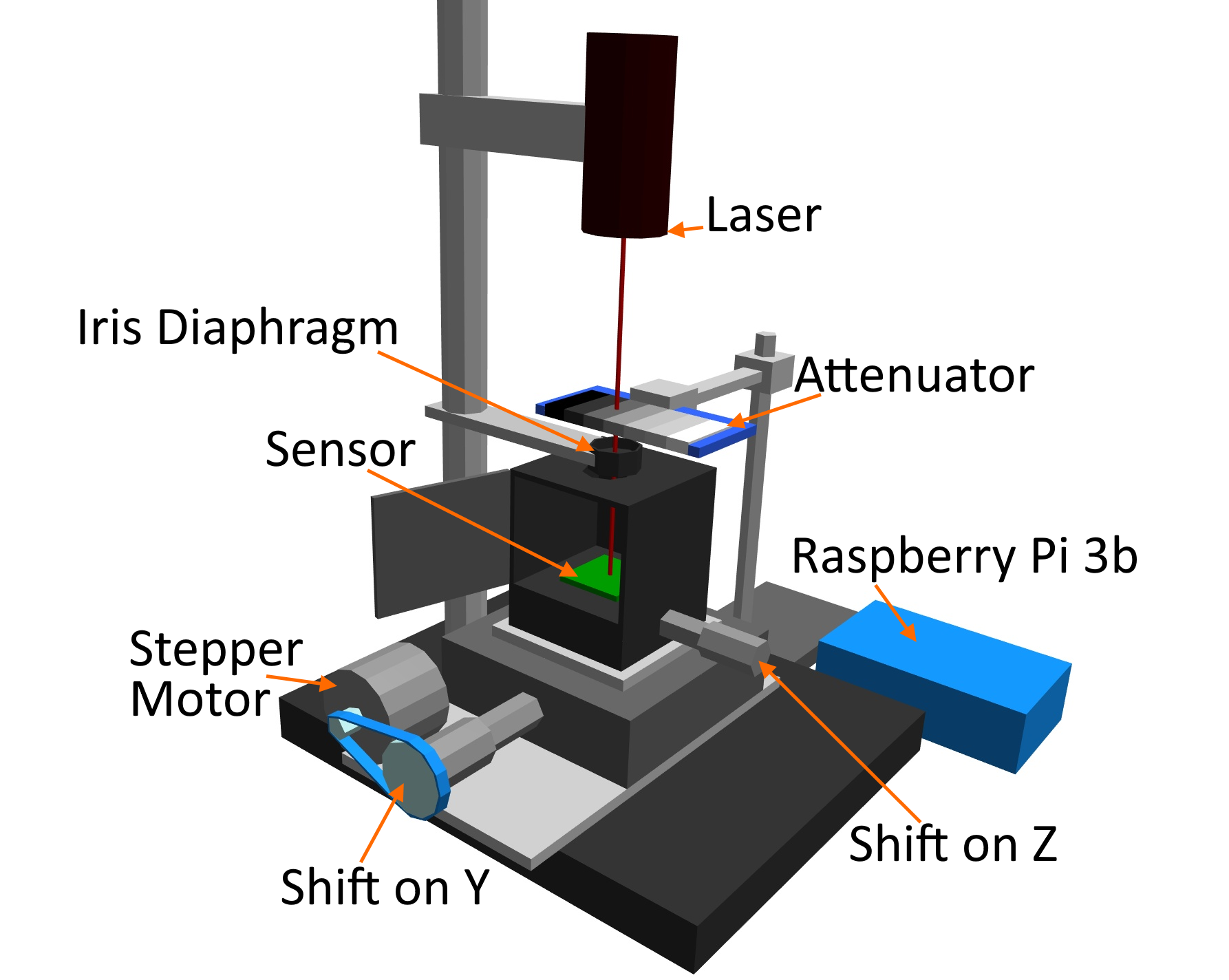}
\caption{Experimental setup to measure the laser spot. The laser remains fixed, while the sensor is moved in the $y$-axis and $z$-axis.} 
\label{fig:assembly}
\end{figure}

To measure the laser beam spot we have designed the experimental setup shown in Fig. \ref{fig:assembly} and tested, applying different methods, three light sensors and two lasers.

At the bottom of the setup there is a micrometer that has a resolution of $0.001$ inches or $25.4$ $\mu$m  connected to a stepper motor which allows a minimum horizontal displacement along the $y$-axis of $10.10 \pm 5.05$ $\mu$m according to the mechanical configuration of the assembly. On top of this micrometer, there is another one that moves the sensor along the $z$-axis, which is controlled manually and has a resolution of $10 \pm 5$ $\mu$m. 
Over the $z$-axis micrometer, each different light sensor (CMOS, photodiode and LDR) was placed. The sensor is partially isolated from ambient light using a dark chamber with a hole in the upper part, which lets the laser pass. Fitting this hole, there is an iris diaphragm \cite{datasheets_Iris_Diaphragms}
to reduce beam aberrations. This iris only lets in the light that passes through the opening, which is regulated approximately to the spot size.

An attenuator (i.e. rectangular step variable metallic neutral density filter, NDL-25S-4 \cite{datasheets_attenuation}) is placed between the laser and the sensor to decrease the intensity of the laser, preventing the saturation of the sensor and keeping the Gaussian profile, especially in the case of the CMOS sensor. The attenuator used has an \textit {optical density} (OD) that varies from $0.1$ to $4$ with a tolerance of $\pm 5\%$, divided into $10$ steps or positions. The OD indicates the attenuation factor provided by an optical filter, i.e. how much it reduces the optical power of an incident beam. OD is related to the transmission, $T$, by $T= 10^{-OD}$. Therefore the transmission varies from 0.794 to 0.0001.

In the upper part of the setup the laser is hold still using a retort stand. The distance between laser and sensor is determined for each measurement and explained in the next Section.

The stepper motor and the sensors were connected to a Raspberry Pi 3b (i.e. a small, low-cost, single-board computer) \cite{datasheets_raspberry3B,raspberry} to control the movement and capture of the sensor data. Python programs were also developed to process the data directly on the Raspberry and to obtain the average results. The scripts can be found in the supplementary material \cite{scripts_data}.

\subsection{Light sensors}

Three different light sensors (CMOS (webcam), photodiode and LDR) were used in this configuration. These sensors are cost-affordable. The webcam can cost circa US\$ 15, while the photodiode can be 5 times more expensive and the LDR less than US\$ 2. 

The first sensor was a 5MP webcam \cite{datasheets_rasp_camera, rasp_camera, git_rasp_camera} which uses CMOS technology. 
We removed the webcam lens to be able to measure the spot size directly, according to the size of the pixels illuminated by the laser, every time a shot was taken. 

The webcam was configured using the picamera package \cite{picamera} for the Raspberry Pi camera module for Python. The settings were:
16 $\mu$s shutter speed (i.e. exposure time) to reduce movement, 2592x1944 image resolution, analog and digital gain at 1 and ISO = 0 to maximally reduce the noise.

The format used for image capture is YUV (Y: intensity, UV color code), which allows images to be captured without loss of detail and with full resolution (1-byte Y value for each pixel).

The second sensor was a single PIN photodiode \cite{datasheets_photodiode}. 
In this case, to determine the beam radius, the photodiode must be moved step-by-step (e.g. similar to the Knife-Edge technique \cite{Wright1992}) perpendicular to the beam from a position where the whole beam hits the sensor until the beam is outside the sensitive area.  This technique will return a Gaussian CDF curve.

The third sensor was an LDR (light dependent resistor) or photo-resistance \cite{datasheets_LDR}. This electronic component decreases the resistance when the light intensity on the sensitive surface increases. Even if it is not a precision device, it is inexpensive. Here we apply the same technique used for the photodiode.

\begin{figure}[H]
\centering
\includegraphics[width=10cm]{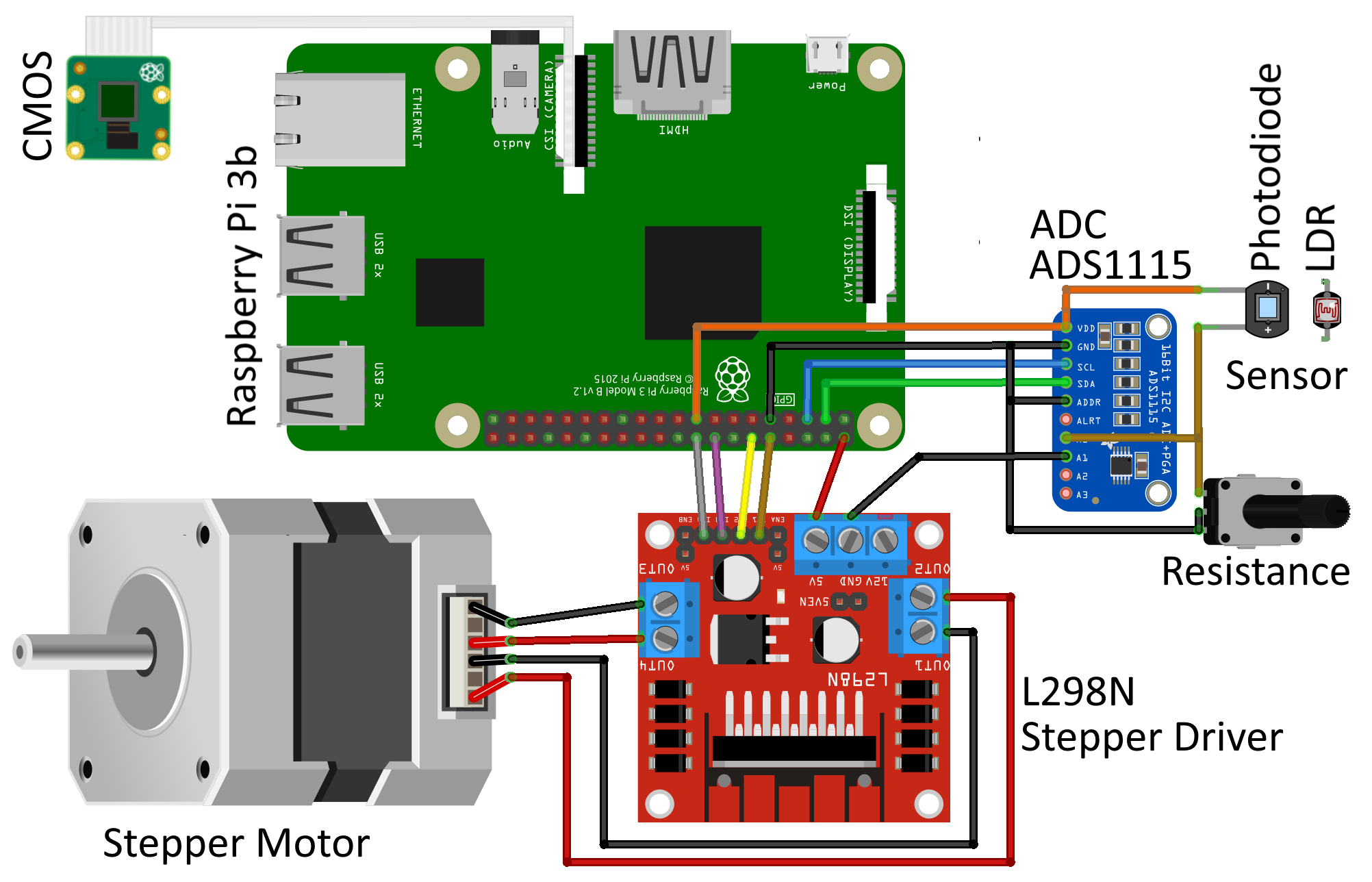}
\caption{CMOS, photodiode and LDR sensors setup.} 
\label{fig:Setup_circuit_all}
\end{figure}

The circuits for the photodiode and the LDR, consist of a variable resistance in series with the sensor. These circuits are connected to an analog-to-digital converter (i.e. ADS1115), which can be read by the Raspberry Pi, as shown in Fig. \ref{fig:Setup_circuit_all}.

\subsection{Lasers}

Two different lasers were used for the measurements to benchmark the performance of the methods with different wavelengths and technologies. One was a generic 5 mW green semiconductor laser that gives a Gaussian spot according to the manufacturer. A model similar to this laser is found in \cite{datasheets_green_Laser}. The second one was a red He-Ne gas laser with power $<4$ mW  \cite{datasheets_Helium-Neon_Laser}.

The laser intensity can be controlled in two ways: for both lasers, attenuators are used, however, for the green laser it is also possible to tune the current to lower its intensity, which is not available for the red laser. 

In addition, the wavelength and power of each laser were measured in the laboratory. The green semiconductor laser had a 532 nm wavelength and a maximum power of 5.57 mW. Once the intensity was regulated with current and attenuators, the power was 2.83 $\mu$W. This laser was attenuated with the density filter reducing its intensity to $0.1$\%  of the incident beam power to avoid saturating the sensor. The red He-Ne gas laser had a 633 nm wavelength and a power of 0.67 mW. 

The green semiconductor laser was used to measure the laser profile and to find the beam minimum waist. Thus, a Gaussian profile was obtained. The red He-Ne gas laser was used to compare the beam radius measurement with the three light sensors. This laser has a very stable spot size with distance. Therefore the beam minimum waist was not  searched for and only the beam radius was measured. In addition, it was necessary to locate the laser at a larger distance than the green laser to avoid saturating the CMOS sensor.  

\section{Laser beam profile measurements}

As a benchmark, the radius and waistline of the green semiconductor laser beam were determined using the CMOS sensor. Next, the beam radius measurements using the red He-Ne gas laser obtained with the CMOS sensor were compared with those of the two other sensors: photodiode and LDR.

\subsection{Radius and waistline estimation}
\label{Laser beam radius measurement}

The initial distance from the green semiconductor laser to the CMOS sensor was determined as follows. First, the distance between laser and sensor, where the area of the beam over the sensor is the smallest (beam waist), was visually searched by moving the laser on the $z$-axis. The distance from the CMOS sensor to the beam waist was measured with a vernier to be $161.94 \pm 0.01$ mm. The beam waist was scanned around this distance in a $\pm$ 5 mm range starting with the minimum distance ($156.94 \pm 0.01$ mm). Then, this minimum distance is considered as the initial position ($z=0$ mm). 

The CMOS sensor was moved away from the laser in $1$ mm steps, obtaining $11$ measurements. In each position, $10$ images or frames were captured and converted to ADC intensity matrices. From these data, the value of the radius in each position was calculated.

To calculate the laser beam radius, the projection of the intensity profile in the $x$ and $y$ axes was made. For each frame, both projections were fitted using the following Gaussian function:
\begin{equation}
g(x)=\frac{a}{\sigma\sqrt{2\pi}}.e^{\frac{-(x - \mu)^2}{2\sigma^2}} + b
\label{eq:gauss_fit}
\end{equation}
obtaining the parameters $a$ (normalization),  
$b$ (background off-set), $\mu$ (mean) and $\sigma$ (standard deviation).

The beam radius for each axis is the 10-frame average of 2$\sigma$. The associated total error was estimated as $\Delta_{tot} = \sqrt {(\sum_{i} {\Delta_{i}}^{2})/N}$, where $N$ is the number of frames and $\Delta_{i}$ is the error for each frame, which is calculated by adding in quadrature the error from the Gaussian fit and the systematic error of 0.5 pixel.

\begin{figure}[H]
\centering
\includegraphics[width=12cm]{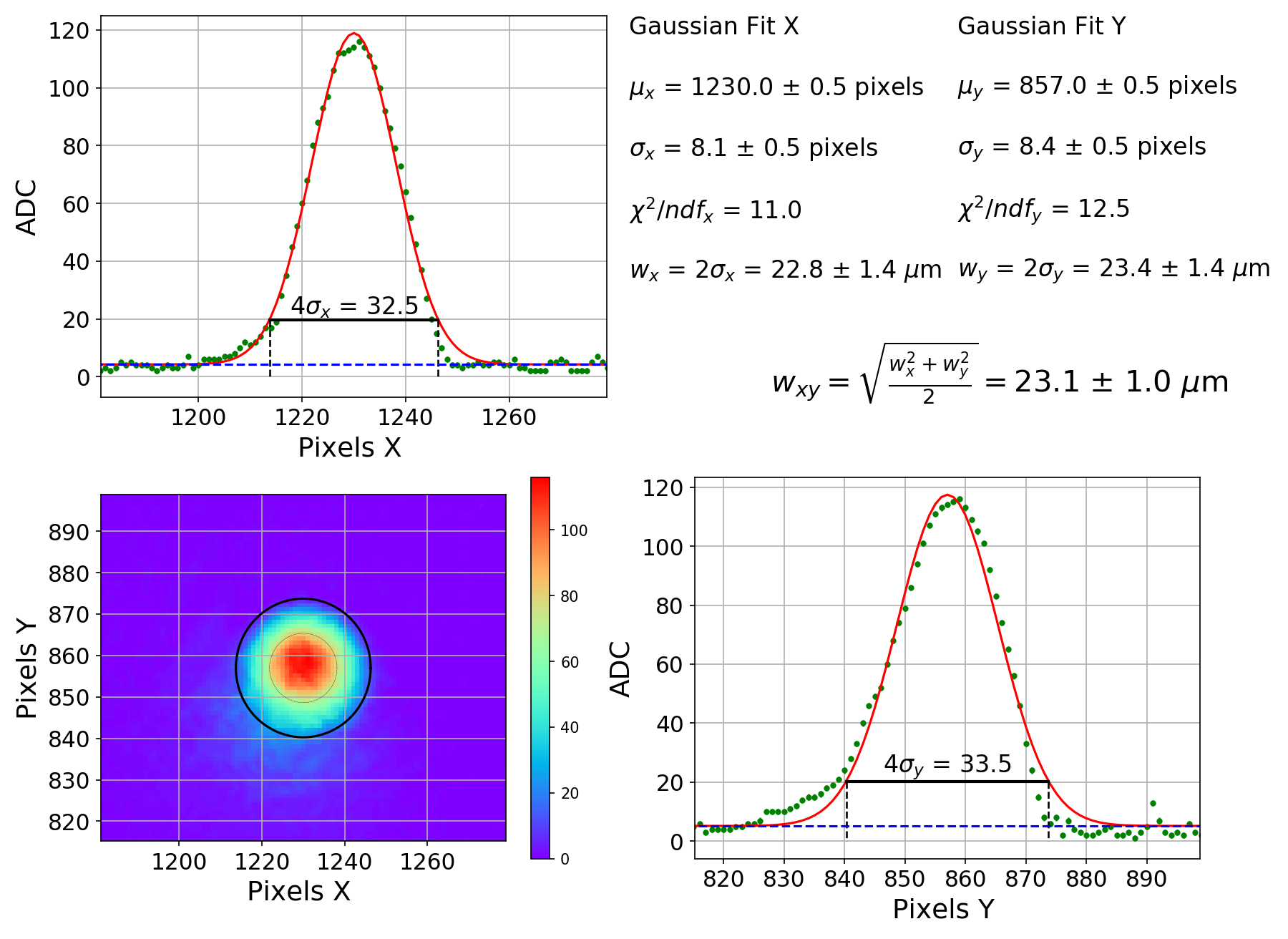}
\caption{Beam radius measurement of the green laser with the CMOS sensor for one frame at 6 mm. Top left and Bottom right: ADC projections along the x-axis and y-axis (in pixels), respectively, of the spot profile captured with the CMOS. The red curves show the Gaussian fits.
Top-right: Gaussian fit parameters for $x$ and $y$ projections (central beam positions $\mu$ and standard deviations $\sigma$ both in pixels, $\chi^{2}$ divided by the number of degrees of freedom ${ndf}$ of the fit), also the radius $w$ in each axis and average spot radius $w_{xy}$ are given in $\mu$m. 
Bottom left: spatial spot profile in pixels measured by the CMOS sensor, the coloured axis represents the ADC counts and the ellipse marks the spot size.}
\label{fig:plot_spot_green}
\end{figure}

To give the results in units of distance, the number of pixels was multiplied by 1.4 $\mu$m, which is the size of each CMOS sensor pixel. In Fig. \ref{fig:plot_spot_green}, the first frame taken with the CMOS sensor at $z=6$ mm is shown, as an example. The projections of the $x$ and $y$ axes of the spot profile are given, as well as the results obtained from the Gaussian fit. 

The average spot radius $w_{xy}$ is calculated by:
\begin{equation}
w_{xy}=\sqrt{\frac{w_x^2+w_y^2}{2}}=\sqrt{\frac{(2\sigma_x)^2+(2\sigma_y)^2}{2}}
\label{eq:rxy}
\end{equation}
where $\sigma_x$ and $\sigma_y$ are the average standard deviation of the Gaussian fit in the x and y axes, respectively.

To determine the laser profile parameters, the least-square fitting of the data at different distances along the beam line was made using Eq. \ref{eq:w_z}. The parameters obtained with the fit are $w_0$, the radius of the minimum possible cross-section, $z_0$, the position of $w_0$ and $z_R$, the Rayleigh range. Using these parameters, the laser wavelength can be determined from Eq. \ref{eq:z_R} and the divergence from Eq. \ref{eq:divrg}.

\begin{figure}[H]
\centering
\subfigure{\includegraphics[width=6cm]{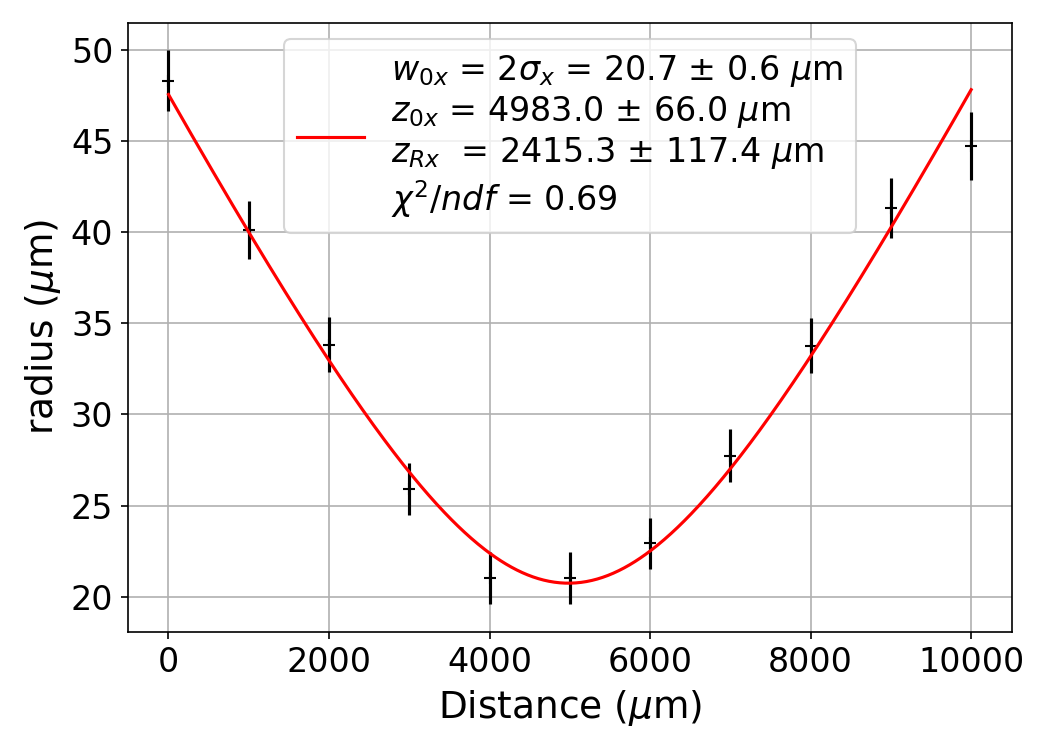}}
\subfigure{\includegraphics[width=6cm]{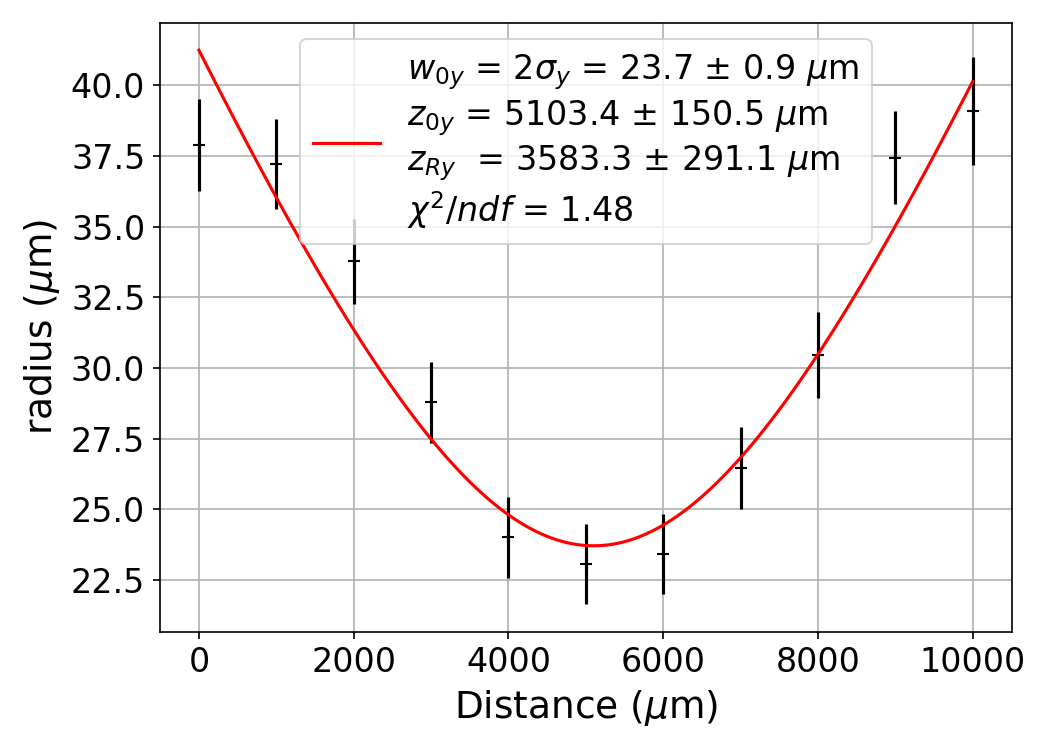}}
\caption{Beam radius as a function of the distance between laser and sensor for the x-axis (Left) and y-axis (Right). The red curves represent the fits using Eq. \ref{eq:w_z}.}
\label{fig:plot_spot_green_x_and_y}
\end{figure}

\begin{figure}[H]
\centering
\includegraphics[width=10cm]{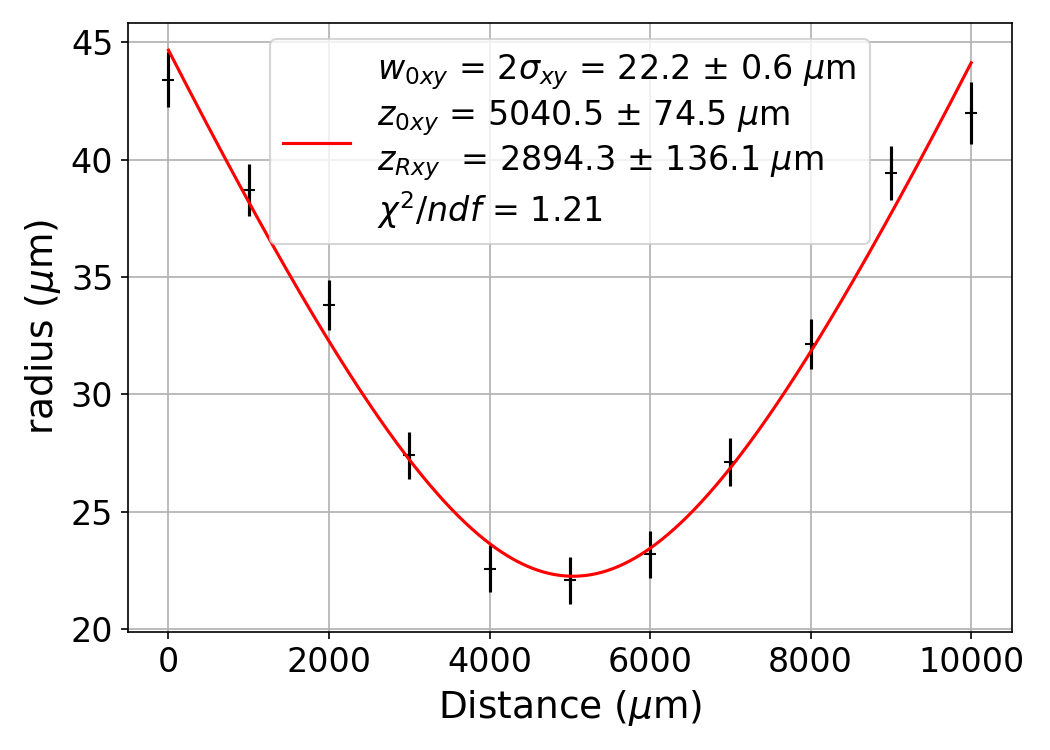}
\caption{Beam average radius as a function of the distance between laser and sensor. The red curve represents the fit using Eq. \ref{eq:w_z}.} 
\label{fig:plot_spot_green_xy}
\end{figure}

The results obtained at different distances from the CMOS sensor are shown in Fig. \ref{fig:plot_spot_green_x_and_y} and Fig. \ref{fig:plot_spot_green_xy}, for the $x$-axis $w_{x}$, $y$-axis $w_{y}$ and average $w_{xy}$ radius, respectively. 

Using the average radius (see Eq. \ref{eq:rxy}), the beam waistline was $w_{0xy} = 22.2 \pm 0.6$  $\mu$m, Rayleigh's range $z_{Rxy} = 2894.3 \pm 136.1$ $\mu$m, the divergence $\theta_d = \frac {w_0}{z_R} = 7.7 \pm 0.4$ mrad and the wavelength  $\lambda = \frac{\pi w_{0}^{2}}{z_R} = 537.0 \pm 38.0$ nm. This estimation of wavelength is in good agreement with an independent measurement in the laboratory, which gave $532$ nm. 

\subsection{Light sensors measurements comparison}

To compare the different sensors we used a red He-Ne gas laser for estimating the beam radius. In this case, the green laser, which has a smaller spot size (between 20 and 45 $\mu$m), was not used since the method applied for the photodiode and LDR has lower resolution than the CMOS sensor. In order to use the photodiode and LDR a step-wise procedure is used, which has a minimum horizontal displacement of 10.1 $\mu$m along the $y$-axis, while the pixel size of the CMOS is 1.4 $\mu$m. Thus, the red laser, which has a larger spot, size was used. For the comparison it is only necessary to measure the spot on one axis, in this case we chose the $y$-axis.

The laser was located $467.15 \pm 0.05$ mm above the sensor for all three sensors to avoid saturation, since this laser's intensity cannot be regulated. In addition, to prevent the saturation of the CMOS, different attenuations were tested. We found that the CMOS did not saturate for transmission of 1\% and 0.1\% of the initial laser intensity. For lower attenuations, the profile cannot be correctly fitted by a Gaussian. The LDR showed smaller variations in the spot size, also decreasing according to the attenuation. Contrarily, we found for the photodiode that the spot size is almost constant for the transmission range between 100\% to 0.1\%. All sensors measurements converge at the attenuator with 0.1\% transmission. All sensors reach their threshold at the last attenuator position, where only 0.01\% of the incident power is transmitted. At this position, the voltage output is compatible with the background. 

\begin{figure}[H]
\centering
\includegraphics[width=10cm]{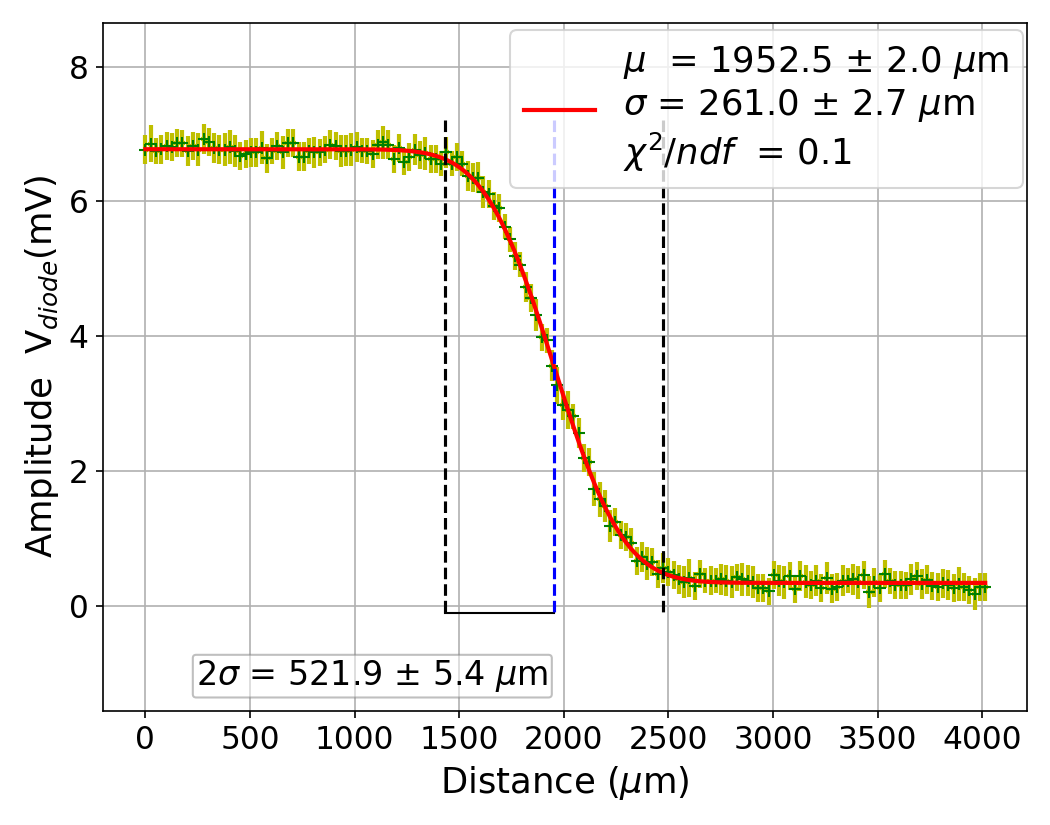}
\caption{Measured photodiode output voltage using a resistance of 2.2 k$\Omega$ and a transmission of $1\%$ of the incident power, as a function of distance from the initial position of the laser beam inside the sensitive area of the sensor 
and a horizontal displacement of 25.4 $\mu$m. 
The red line is the Gaussian CDF fit of the data from which the spot radius is obtained. The blue dashed line shows the position where half of the beam spot is inside the sensitive area of the sensor. The distance between the blue line and any of the black dashed lines represents the spot radius ($w_y=2\sigma$).} 
\label{fig:plot_spot_diode_R2.2k_red_y}
\end{figure}

The laser spot was measured with the photodiode, using the method explained in Sec. \ref{sec:expt_setup}, regulating the resistance of the circuit (see Fig. \ref{fig:Setup_circuit_all}) to improve the sensitivity. Each set of voltage measurements began with the laser pointing inside the sensor. Then, the sensor was moved stepwise along the $y$-axis until the laser was outside the sensitive area of the sensor. In each step, 10 measurements were made. The resulting data was the average. The points obtained from these measurements follow a Gaussian CDF, as seen in Fig. \ref{fig:plot_spot_diode_R2.2k_red_y}, since the profile of the laser is Gaussian.

Thus, we fit these data using the corresponding following function:
\begin{equation}
f(x)=\frac{a}{2} \left [1+\mathrm{erf} \left ( \frac{x-\mu}{\sigma\sqrt{2}} \right )\right ] + b
\label{eq:gauss_fit_cdf}
\end{equation}
where the parameters to fit are: $a$ the normalization, $\mu$, the Gaussian mean (position where half of the beam spot is inside the sensitive area of the detector), $\sigma$, the standard deviation ($w_y$=2$\sigma$) and $b$ the off-set from the background. 

Fig. \ref{fig:plot_spot_diode_R2.2k_red_y} shows an example of the Gaussian CDF fit and the obtained parameters using the photodiode data with an attenuator that allows a transmission of $1\%$ of the incident power and resistance of $2.2$ k$\Omega$. A different value of the resistance would not affect the spot size, it would only change the error since the amplitude is changed.

In addition, the step distance was varied from 10 $\mu$m to 55 $\mu$m using a fixed attenuation. 
The beam radii obtained along the $y$-axis is stable, since the spot is large enough, it does not depend, in this distance range, on the precision of the steps.

Then the LDR sensor was used to measure the laser spot with an 11.5 $\Omega$ resistance, following the same method that was used for the photodiode. Different measurements were performed varying the series resistance of the circuit. And it was observed that the results obtained with the LDR depend on the resistance of the circuit, varying the sensitivity of the sensor. This does not guarantee a reliable measurement of the spot. Only a rough estimate can be given. This is because the LDR does not have a continuous surface that measures the intensity of the beam, rather it has a zigzag of the light-sensitive semiconductor material with active and non-active areas. Also, the LDR own resistance can vary up to three orders of magnitude due to changes in the intensity of the laser. We estimate that these factors account for an approximately 10\% error.

Finally, the laser spot was measured using the CMOS sensor following the same method described in Sec. \ref{Laser beam radius measurement}. To compare the three sensors results, the CMOS values are taken as the standard for comparison. The errors are calculated using Eqs. \ref{eq:err_2} and \ref{eq:n_sigm}, where $\sigma$ is the combined error of $\Delta w_{sensor}$ the error in the beam radius estimation obtained for either photodiode or LDR and $\Delta w_{CMOS}$ the error of the CMOS. The difference between measurements of different sensors will be given in terms of the number of standard deviations, $n_\sigma$. 

\begin{equation}
\sigma =\sqrt{(\Delta w_{CMOS})^{2} + (\Delta w_{sensor})^{2} }
\label{eq:err_2}
\end{equation}

\begin{equation}
n_\sigma=\frac{\left | w_{CMOS} - w_{sensor} \right |}{\sigma}
\label{eq:n_sigm}
\end{equation}

The errors include the statistical and systematic uncertainties, which are for the CMOS half the pixel size, that is 0.7 $\mu$m, and for the photodiode and LDR the horizontal displacement of 5 $\mu$m. The systematic error for the LDR also adds the aforementioned factors by which the spot cannot be accurately measured, such as the variation of resistance due to the effects of intensity, resistance tolerance, hysteresis effects, etc.. These is estimated to be approximately 10\% of measurement.

\begin{figure}[H]
\centering
\includegraphics[width=12cm]{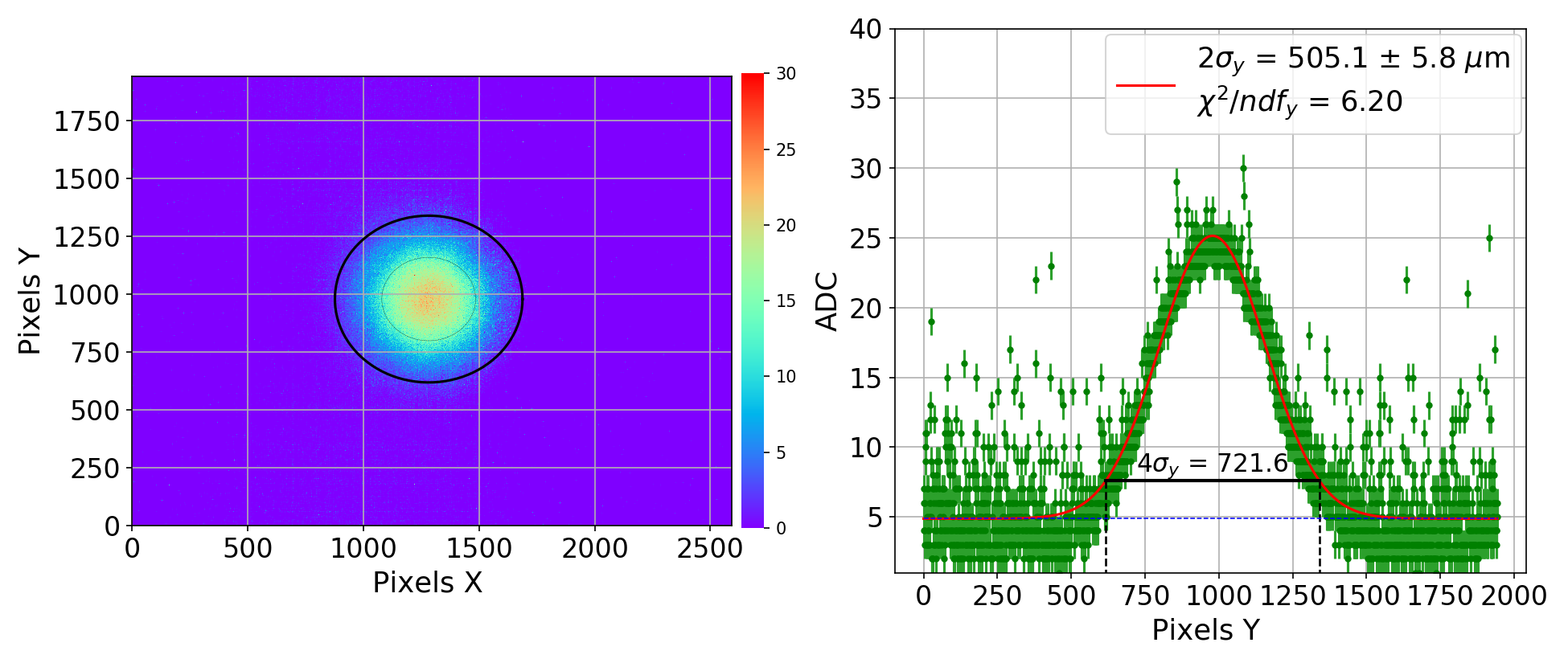}
\subfigure{\includegraphics[width=6.cm]{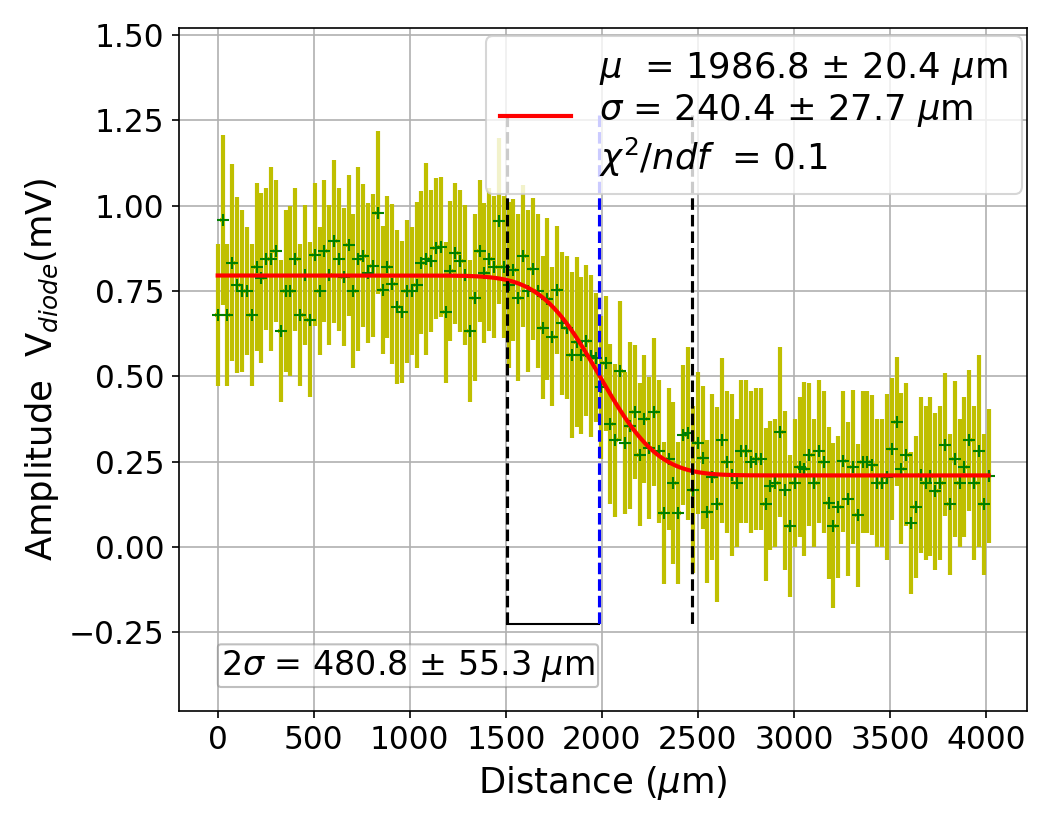}}
\subfigure{\includegraphics[width=6.cm]{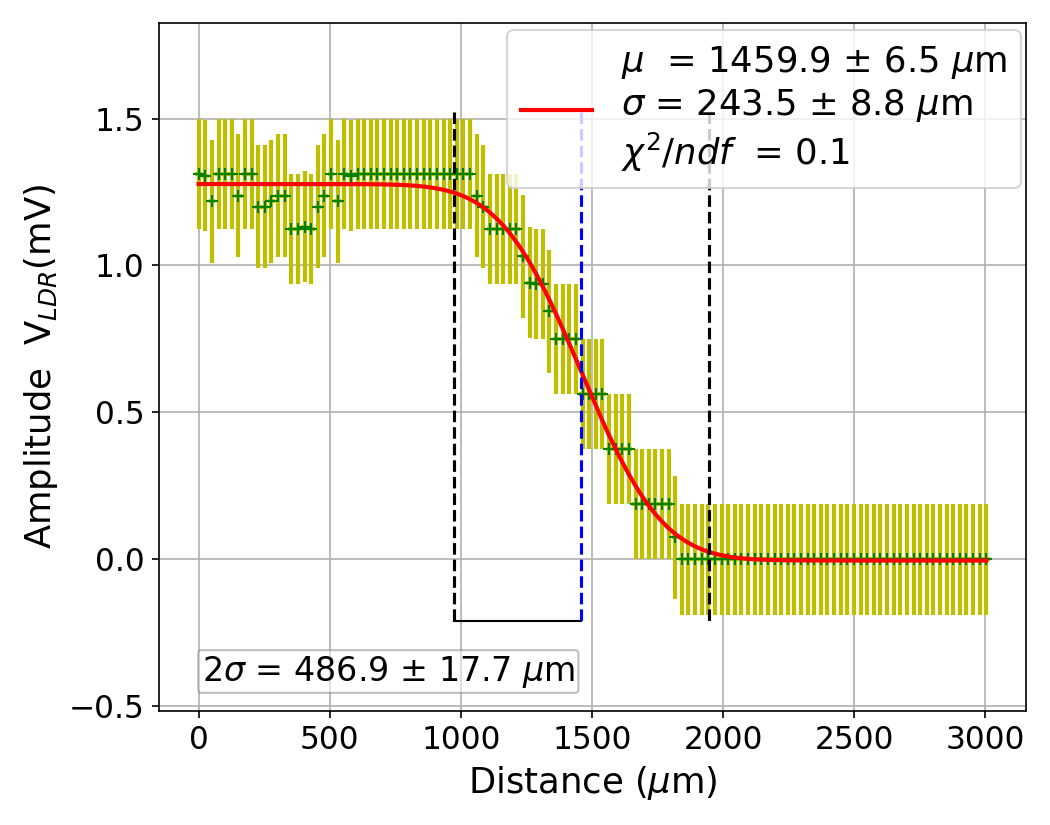}}
\caption{Spot estimation with the different sensors for the attenuator with 0.1\% transmission. Top left: spatial spot profile in pixels measured by the CMOS sensor (one frame), the colored axis represents the ADC counts and the ellipse marks the spot size. Top right: ADC projections along the y-axis (in pixels), of the spot profile captured with the CMOS (one frame). The red curve shows the Gaussian fits. Bottom left/right: photodiode/LDR output voltage as a function of distance (horizontal displacement of 25.4 $\mu$m). The red lines are the Gaussian CDF fit of the data from which the spot radii are obtained.} 
\label{fig:fit_CMOS_diode_LDR}
\end{figure}

Table \ref{table:Comp_CMOS_diode_LDR} summarizes the results of all three sensors at 0.1\% transmission: CMOS, photodiode and LDR. Considering the uncertainties, the values obtained with the photodiode and LDR are compatible with that of the CMOS (webcam), less than 0.5 $\sigma$ difference. Fig \ref{fig:fit_CMOS_diode_LDR} shows the corresponding fits for all sensors.

\begin{table}[H]
\centering
\begin{tabular}{l|l|l|}
\cline{2-3}
& $r_{sensor} (\mu m)$ & $n_\sigma$ \\ \hline
\multicolumn{1}{|l|}{CMOS}              & 500.4 $\pm$ 5.5      &            \\ \hline
\multicolumn{1}{|l|}{Photodiode}    & 480.8 $\pm$ 56.7   & 0.4   \\ \hline
\multicolumn{1}{|l|}{LDR}          & 486.9 $\pm$ 54.1      & 0.3    \\ \hline
\end{tabular}
\caption{Comparison of the spot radius results obtained with the CMOS with those of the photodiode and LDR at 0.1\% transmission.} 
\label{table:Comp_CMOS_diode_LDR}
\end{table}

The measurement of the spot with the photodiode can be improved using less attenuation or the average of the measurements with the different attenuators, since it was observed that the spot size is almost constant for the transmission range between 100 \% to 0.1 \%. For example, if we use the 10\% transmission attenuator, the spot value would be $506.4 \pm 5.3$  $\mu$m.

\section{Discussion}
\label{sec:discussion}

We have designed and tested the performance of an automatic setup for measuring the spot radius of laser beams using three different cost-affordable light sensors. Two common profiling techniques were implemented: imaging for the CMOS 2D array and scanning knife-edge-like for the single photodiode and LDR.

The simplest and with the highest precision method for beam profiling was using the CMOS sensor. The error in determining the minimum spot radius is less than 3\%, while the error on the waistline position is 1.5\%. As a separate cross-check of the method, the wavelength of the laser could be determined with a 1\% error. This shows that it is possible to use a low-cost webcam to determine the profile of a laser with good precision. 

Our results have smaller uncertainties than those obtained in \cite{Purvis20191977} where the error of the beam waist is 26\% with a CMOS sensor compared to 1\% in our case. For the method that uses a photodiode our errors are similar around 12\% for the highest attenuation. However, for a different attenuation our error can improve to less than 2\%.

Nevertheless, this method has also some limitations. Due to the CMOS greater sensitivity, the sensor can be saturated and it can produce blooming with a high intensity laser, distorting its real profile. This problem can be overcome by using attenuators. 

The photodiode using the knife-edge-like technique is the most stable method. It is not affected by the change in laser intensity until the attenuation is very high. Its precision for the spot radius measurement is better than 2\% under standard conditions. However, a setup with a stepper that moves the sensor along one axis is needed and the price of the photodiode is 5 times more expensive than the CMOS webcam.

The LDR measurement has a strong dependence on the series resistance of the circuit. In addition, the LDR own resistance varies about three orders of magnitude depending on the light intensity. It has also a slow response and does not provide linearity with lighting. This limits its precision to accurately measuring the spot of a laser, having an estimated error of 11\%. However, it has a high sensitivity and given its considerable low cost, it can be used to have a rough estimate of the spot value.

A limitation for the smallest spot that can be measured with the knife-edge-like technique of the photodiode and LDR measurements is related to the mechanical system for horizontal displacement. Currently, the smallest step is 10.1 $\mu$m, while the CMOS grid has a pixel size of 1.4 $\mu$m. If the mechanical system is improved, smaller spots could be measured with the photodiode and LDR.

Furthermore, the data collection system for spot determination can be improved by automating the movement in the $z$-axis. This would increase the speed and precision of the measurement. 

\section*{Acknowledgements}

The authors gratefully acknowledge the Direcci\'on de Gesti\'on de la Investigaci\'on (DGI-PUCP) for funding under Grant No. DGI-2019-3-0044. C.S. acknowledges support from the Peruvian National Council for Science, Technology and Technological Innovation scholarship under grant 236-2015-FONDECyT.

We would also wish to thank R. Sanchez from the Applied Optics Group and the Quantum Optics Group for letting us use their lasers and equipment, as well as Y. Coello for the photodiode. We also thank J. A. Guerra for useful discussions and suggestions.

\bibliography{mybibfile}

\end{document}